\documentclass[preprint,prd,noshowpacs]{revtex4}
\usepackage{amssymb}
\usepackage{amsmath}

\begin{document}

\title{Entropic derivation of $F=ma$ for circular motion}

\author{Michael Duncan}
\email{duncanmi@mail.fresnostate.edu}
\affiliation{Department of Physics, CSU Fresno, Fresno, CA 93740 USA}

\author{Ratbay Myrzakulov}
\email{cnlpmyra1954@yahoo.com}

\author{Douglas Singleton}
\email{dougs@csufresno.edu}
\affiliation{Department of Physics, CSU Fresno, Fresno, CA 93740 USA}

\affiliation{Eurasian International Center for Theoretical Physics, Eurasian National University, 
Astana 010008, Kazakhstan}

\date{\today}

\begin{abstract}
We examine the entropic picture of Newton's second law for the case of circular motion. It is shown that one must 
make modifications to the derivation of $F=ma$ due to a change in the effective Unruh temperature 
for circular motion. These modifications present a challenge to the entropic derivation of Newton's 
second law, but also open up the possibility to experimentally test and constrain this model 
for large centripetal accelerations.     
\end{abstract}

\maketitle

The idea that there is a connection between black holes and thermodynamics goes back almost forty years 
\cite{bekenstein, bardeen}. More recent work has proposed an even closer connection between
gravity and thermodynamics. In \cite{jacobson} the Einstein field equations were arrived at using thermodynamic 
arguments. In \cite{pad, verlinde, banerjee, mureika} general proposals were given of gravity as an emergent phenomenon
using an analogy to thermodynamics entropy. 

The holographic, entropic model of gravity proposed in \cite{verlinde} has attracted a lot of attention.
This model has given new perspectives on cosmological models -- it has been used to 
derive the Friedman equations \cite{cao} and to give an explanation of early time \cite{cai} and late time 
acceleration \cite{cai2, easson} of the expansion rate of the Universe. 

Further constraints on and insights into the entropic model proposed in \cite{verlinde} have been
made by various authors. The paper \cite{nicolini} examined the combination of entropic gravity and 
non-commutative space-times to show that one could still arrive at an entropic derivation of Newton's laws
under this extension. The article \cite{mann} placed bounds on possible photon and graviton masses in
the context of gravity as an entropic phenomenon. Cosmological constraints on the entropic gravity model
were investigated in \cite{wei}. Finally, the article \cite{kobakhidze} gave arguments that the gravity as
entropy model might be ruled out by experiments with ultra-cold neutrons in the gravitational field of Earth. 

In this letter we examine Verlinde's \cite{verlinde} entropic derivation of $F=ma$ for circular motion and show
that significant modifications must be made to preserve the derivation of $F=ma$. 
These modifications might rule out or give experimental constraints to the entropic picture of $F=ma$. 
To begin we briefly review the entropic derivation of $F=ma$ for linear motion to set up the framework and
notation for studying the case of circular motion. To begin one assumes that 
a particle of mass $m$ changes its position by $\Delta x$ with respect to some holographic screen 
by one Compton wavelength,
\begin{equation}
\label{DX}
\Delta x = \frac{\hbar}{mc} ~.
\end{equation}
For the case of linear acceleration it is natural to equate the holographic screen with
the Rindler horizon \cite{culetu}. Then with the change in position, $\Delta x$, one associates a 
change in entropy given by 
\begin{equation}
\label{DS}
\Delta S = 2 \pi k _B ~,
\end{equation}
where $k_B$ is Boltzmann's constant. The factor of $2 \pi$ is specifically chosen to cancel a factor of $2 \pi$ coming from
the Unruh temperature as we will see below. Using the equation for $\Delta x$ to insert $1$ in the change in entropy 
equation we arrive at
\begin{equation}
\label{DS1}
\Delta S = 2 \pi k_B \frac{mc \Delta x}{\hbar} ~.
\end{equation}
From arguments about the forces on polymers in a heat bath one can motivate the following relationship
between force, $F$, temperature, $T$, change in entropy, $\Delta S$ and displacement or stretching of 
the polymer, $\Delta x$
\begin{equation}
\label{force-s}
\frac{F}{T} = \frac{\partial S}{\partial x} \rightarrow F \Delta x = T \Delta S ~.
\end{equation}
At this point one introduces the Unruh temperature \cite{unruh} for $T$ in
\eqref{force-s}. The Unruh temperature is due to the thermal heat bath seen by an ever accelerating observer
who has acceleration $a$. Its explicit value is 
\begin{equation}
\label{unruh-temp}
k _B T_L = \frac{1}{2 \pi} \frac{\hbar a}{c} ~.
\end{equation}
The subscript $L$ stands for linear acceleration. This association of the temperature in
\eqref{force-s} with the Unruh temperature in \eqref{unruh-temp} is the crucial step in arriving
at Newton's Second Law. As we will show below, for circular motion, one has a different temperature
and thus one must modify the derivation of $F=ma$ for this case. 

From \eqref{unruh-temp} one now sees the reason for the factor of $2 \pi$ in the equation for $\Delta S$. 
Putting the results of \eqref{force-s} and \eqref{unruh-temp} together yields Newton's Second Law
$$
F=ma
$$
In the case of linear acceleration the holographic screen, with respect to which the change in entropy $\Delta S = 2 \pi k _B$
occurs, is identified with the Rindler horizon \cite{culetu} of the accelerating observer. 

For circular motion the effective Unruh temperature is not given by \eqref{unruh-temp}. We recall some keys points
leading to \eqref{unruh-temp} for linear acceleration in order to see the differences 
that arise with the entropic derivation of $F=ma$ in the case of circular motion. We begin
by considering an Unruh-Dewitt detector -- a system with two energy levels $E_0 < E$ -- which moves 
through space-time. If, for a given space-time path in some particular space-time background, such a detector
gets excited to the higher energy level, $E$, from the lower level, $E_0$, this is an indication of Hawking-like
or Unruh-like radiation. For simplicity we consider an Unruh-Dewitt detector which is coupled to a scalar field, $\phi (x)$,
via a monopole coupling. The interaction for this system is $g \mu (x) \phi (x)$ -- where $g$ is the coupling constant, $\mu (x)$, 
is the detector's monopole moment, $\phi (x)$ is the scalar field, and $x=x^\mu (t)$ is the detector's space-time path . 

For linear acceleration one can calculate \cite{birrell} the ratio of 
the population of electrons in the excited state, $E$, versus the ground state, $E_0$. This
ratio also equals the ratio of transition probabilities per unit 
time to excite $E_0 \rightarrow E$ versus to de-excite $E \rightarrow E_0$. The result is 
\begin{equation}
\label{linear-pop}
\frac{P_{excite}}{P_{de-excite}} = \exp \left( {-\frac{2 \pi c \Delta E}{a \hbar}} \right)~.
\end{equation}
Equating \eqref{linear-pop} with a Boltzmann distribution, $\exp (- \Delta E / k_B T )$,
gives the Unruh temperature for linear acceleration \eqref{unruh-temp}. 

A similar analysis for circular motion \cite{bell, sing2} does not give such a simple
result as in \eqref{linear-pop}. In the limit $\Delta E \gg \hbar a / c$ one finds the ratio
of the population of electrons in $E$ versus $E_0$ is
\begin{equation}
\label{circle-pop}
\frac{P_{excite}}{P_{de-excite}} \approx \frac{a \hbar}{4 \sqrt{3} c \Delta E} \exp 
\left( {-\frac{2 \sqrt{3} c \Delta E}{a \hbar}} \right)~,
\end{equation}
instead of \eqref{linear-pop}. Already at this point we want to emphasize the difference between the
two expression \eqref{linear-pop} for linear acceleration and \eqref{circle-pop} for circular acceleration. 
First, while \eqref{linear-pop} is exact the result for circular motion \eqref{circle-pop} requires the
limit $\Delta E \gg \hbar a / c$. Second, the circular result \eqref{circle-pop} has a pre-factor in front to
the exponential which depends on the variables $\Delta E$ and $a$. The net result, as we detail below, is that
for circular motion the spectrum is not thermal; it is only approximately thermal in the limit
$\Delta E \gg \hbar a / c$ such that the exponential dominates the pre-factor. Assuming this limit one can read off 
an approximate temperature from the exponential part of \eqref{circle-pop} which gives a circular Unruh 
temperature \cite{bell, sing2} of 
\begin{equation}
\label{unruh-temp2}
k _B T_C = \frac{1}{2 \sqrt{3}} \frac{\hbar a}{c}~,
\end{equation}
where the subscript $C$ stands for circular. (The connection between the results of \cite{bell} and
the entropic picture of gravity of \cite{verlinde} has also been noticed in \cite{kopp}). This temperature 
\eqref{unruh-temp2} is larger than the one for linear acceleration \eqref{unruh-temp}. If one retains everything as before 
in the derivation of $F=ma$ -- equations \eqref{DX} through \eqref{force-s} -- but now uses
\eqref{unruh-temp2} instead of \eqref{unruh-temp} one ends up with following relationship between
force and acceleration
\begin{equation}
\label{f=ma}
F = \frac{\pi}{\sqrt{3}} ma = \frac{T_C}{T_L} ma
\end{equation}
This is clearly wrong since $F=ma$ applies to circular motion. If \eqref{f=ma} were correct then this
would have already been noticed in any of a number of classical, Newtonian examples of circular motion. 
In order to correct the entropic derivation for circular motion one must take a different relationship 
between $\Delta S$ and $k_B$ from that given in \eqref{DS}. One needs to to replace \eqref{DS} with
\begin{equation}
\label{DS2}
\Delta S_C = 2 \sqrt{3} k_B ~.
\end{equation}
The subscript $C$ stands for circular motion. It may seem {\it ad hoc} to propose two separate 
relationships -- \eqref{DS} and \eqref{DS2} -- for linear and circular motion, but this is a direct consequence of
having different temperatures -- \eqref{unruh-temp} and \eqref{unruh-temp2} -- for linear and circular motion.

There is a further complication. In comparing the circular case \eqref{circle-pop} with the linear case
\eqref{linear-pop} one notices that the former expression has a pre-factor which depends on $\Delta E$ and
$a$. This is an indication that the spectrum for the circular Unruh effect if not exactly thermal. The technical reason
for the non-thermal nature of the circular Unruh spectrum comes from the fact that the Wightman function for 
circular motion has a $\sin ^2 (...)$ in the denominator in place of the $\sinh ^2 (...)$ that occurs for the linear 
acceleration case. As a result the spectrum is not exactly thermal and one can not do the contour integration analytically as is possible
for linear acceleration (section 4.4 of \cite{birrell} gives details of the contour integration for the linear case). 
For circular motion one must do the contour integration numerically \cite{muller}. From this numerical evaluation one
finds that the spectrum for circular Unruh radiation is only approximately thermal. The spectrum is closer to thermal 
the more that the limit $\Delta E \gg \hbar a / c$ (which is required for the validity of \eqref{circle-pop}) is satisfied.
The thermal nature of the linear acceleration case versus the approximately thermal nature of the circular acceleration case
shows that one can not think of the circular Unruh effect as the linear Unruh effect plus some additional effect due to
the kinematics of circular motion.

To see the effect that the non-thermal nature has on the arguments from \eqref{unruh-temp2} to \eqref{DS2} (which 
essentially assumes a thermal spectrum for the circular Unruh effect) one can define an effective temperature \cite{bell}
\begin{equation}
\label{t-eff}
k_B T_{C(eff)} = -\frac{\Delta E}{\ln \left(  \frac{P_{excite}}{P_{de-excite}}\right)} = \frac{\Delta E}{R + \ln (2 R)} ~,
\end{equation}
where $R \equiv (2 \sqrt{3} c \Delta E) / ( a \hbar )$. This variable satisfies $R \gg 1$ since $\Delta E \gg \hbar a / c$. 
Taking the ratio of \eqref{t-eff} to \eqref{unruh-temp} gives
\begin{equation}
\label{t-ratio}
\frac{T_{C(eff)}}{T_L} = \frac{\pi}{\sqrt{3}} ~ \frac{R}{(R + \ln (2 R))} ~.
\end{equation}
For $R \gg 1$ this ratio is close to the simple ratio ($\pi/\sqrt{3}$) given in \eqref{f=ma}. Even in case 
when one takes $R =10$ (so that $R \gg 1$ is no really valid) one finds a ratio $T_{C(eff)} / T_L \approx 1.4$ which
would give $F= 1.4 ma$ for circular motion. This again would be in disagreement with a host of classical, Newtonian examples
of circular motion which require $F=ma$. The remedy for this would be to further modify the relationship
between $\Delta S$ and $k_B$ in the case of circular motion. For general $R$ one should define $\Delta S$ as
\begin{equation}
\label{DS3}
\Delta S_C = 2 \sqrt{3} \left( 1 + \frac{\ln (2R)}{R} \right) k_B ~.
\end{equation}
This relationship depends on $R$ i.e. on $a$ and/or $\Delta E$. This is even more complex and contrived than 
needing to have a different relationship between $\Delta S$ and $k_B$ for linear \eqref{DS} and
circular acceleration \eqref{DS2}. Now one needs a relationship between $\Delta S$ and $k_B$ which 
depends on $R$. 

Having a relationship between $\Delta S$ and $k_B$, like that in \eqref{DS3}, which depends on 
$R =(2 \sqrt{3} c \Delta E) / ( a \hbar )$, would make the entropic model of \cite{verlinde} 
too contrived. However, one could propose, that for circular motion, one 
use the $R$-independent relationship of \eqref{DS2}. Then the $\ln (2 R) /R$ term in \eqref{DS3} 
would be a small (since $R \gg 1$) correction term to $\Delta S$. Then since $F \propto \Delta S$ 
this would provide a small correction to $F=ma$ for circular motion. For example, with
$R = 10^6$ one has $\ln (2 R) /R \approx 1.45 \times 10 ^{-5}$ which would give a small but possibly 
measurable effect for circular motion. However, to achieve an $R=10^6$ for an energy splitting of
$\Delta E = 1$ eV would require an enormous acceleration -- $a = 2.5 \times 10^{17} ~ m/s^2$. Such
accelerations (or greater) can be obtained in particle accelerator storage rings \cite{bell}. It might also be 
possible to test this for electrons in atoms -- for an electron orbiting a proton at a distance of $10^{-10}$ m 
a classical estimate of the acceleration of the electron would give $2.5 \times 10 ^{22} ~ m/s^2$. 
In fact such considerations might already rule out the entropic model of $F=ma$ if one accepts the assumptions
made at the beginning of this paragraph.  

In this letter we point out that the proposal \cite{verlinde} to derive $F=ma$ from entropic arguments
would need to be significantly modified for circular motion. One would need to modify the
relationship between $\Delta S$ and $k_B$ given in \eqref{DS} which works for linear acceleration, but
not for circular acceleration. One final point worth noting is that for circular motion there is no horizon as in the 
case of linear acceleration where one has the Rindler horizon. For the linearly accelerating case it was suggested
\cite{culetu} that the holographic screen, with respect to which $\Delta x$ from \eqref{DX} is defined,
be identified with the Rindler horizon. For circular motion there is no horizon -- since the observer is circling 
around a fixed point there is no region of the space-time from which photons will not be able to reach the
observer. However there is still a special surface that can be defined 
which could be identified with the holographic screen -- the light surface \cite{ohanian}. To illustrate this
we first consider a rotating frame i.e. an observer fixed at some point taken to be the origin and rotating around this 
point with angular velocity $\omega$. The metric seen by this observer is obtained by applying the transformation
$t'=t$, $z'=z$, $r'=r$ and $\varphi ' = \varphi - \omega t$ to Minkowski space-time in cylindrical coordinates
$ds^2 =dt^2 - dz^2 - dr^2 - r^2 d \varphi ^2$ yielding
\begin{equation}
\label{rotating}
ds^2 = (1 -\omega ^2 r^2 ) dt^2 - dz^2 - dr^2 - 2 \omega r^2 dt d \varphi - r^2 d \varphi ^2 ~,
\end{equation}
where we have relabeled the coordinates. The light surface is defined by $\omega r =1$ (we are using $c=1$ here). Physically this
corresponds to the distance from the origin where the linear velocity associated angular velocity, $\omega$, is equal to 
the speed of light. Although the light surface is a different concept from a horizon,
it nevertheless picks out some special surface which one might identify the holographic screen. For an orbiting observer
(i.e. one undergoing circular motion not simply rotating around a fixed point) one should simply displace the
coordinates of the rotating system \eqref{rotating} from the origin by $R$. In this case $R$ would be the radius
of the circular orbit. In this case the light surface would be defined by $\omega \sqrt{r^2 +R^2 + 2Rr \cos \varphi} =1$
where $\varphi$ is the angle between orbit radius $R$ and the position $r$. In either the rotating or orbiting cases the 
light surface, while much different from the Rindler horizon of a linearly accelerating observer, presents a natural 
surface with which to associate the holographic screen. 

In summary circular motion appears to present a challenge to the entropic picture of Newton's Second law.

\end{document}